\documentclass[11pt]{article} 
\usepackage{latexsym}
\font\thinlinefont=cmr5
\input{pictex.sty}

\newcommand{\UGR}{\em Dpto. de F\'{\i}sica Te\'orica y del Cosmos,  
		       Universidad de Granada, 18071 Granada, Spain}   
\newcommand{\UP}{\em Dip. di Fisica, 
		       Universit\'a di Padova, 35131 Padova, Italy}   
\newcommand{\UKA}{\em Institut f\"ur Teoretische Physik, Universit\"at
                      Karlsruhe, 76128 Karlsruhe, FR Germany}

\newcommand{\be}{\begin{equation}}
\newcommand{\ee}{\end{equation}}
\newcommand{\bea}{\begin{eqnarray}}
\newcommand{\eea}{\end{eqnarray}}
\newcommand{\ba}{\begin{array}}
\newcommand{\ea}{\end{array}}
\newcommand{\bfi}[1]{\begin{figure}[#1]}
\newcommand{\efi}{\end{figure}}
\newcommand{\bpi}[2]{\begin{picture}(#1,#2)}
\newcommand{\epi}{\end{picture}}

\newcommand{\nn}{\nonumber\\}

\def\ie{{\it i.e.\/}}

\newcommand{\g}{\gamma}
\newcommand{\prop}{\Delta}
\newcommand{\propm}{{\Delta}_m}
\newcommand{\dsl}{\not \! \partial}
\renewcommand{\d}{\partial}
\newcommand{\T}{\mathrm{T}}
\newcommand{\B}{\mathrm{B}}
\newcommand{\A}{\mathrm{A}}
\renewcommand{\O}{\mathcal{O}}

\def\QQa{\renewcommand{\baselinestretch}{1.3}\Huge\large\normalsize}

\makeatletter 
\def\secteqno{\@addtoreset{equation}{section}% 
\def\theequation{\thesection.\arabic{equation}}} 
 
\def\endsecteqno{\def{theequation\{\@ifundefined{chapter}% 
{\arabic{equation}}{\thechapter.\arabic{equation}}}} 
\makeatother

\begin{document}

%%%%%%%%%%%%%%%%%%%%%%%%%%%%%%%%%%%%%%%%%%%%%%%%%%%%%%%%%%% 
%        Titulo                                    % 
%%%%%%%%%%%%%%%%%%%%%%%%%%%%%%%%%%%%%%%%%%%%%%%%%%%%%%%%%%% 
\pagestyle{empty} 
{\hfill \parbox{6cm}{\begin{center} 
				    UG-FT-73/97 \\ 
				    KA-TP-10-1997 \\
				    DFPD 97/TH 38 \\
				    September 1997 
		      \end{center}}} 
	      
\vspace*{2cm}                               
\begin{center} 
\large{\bf Constraining differential renormalization \\ }
\large{\bf in abelian gauge theories} 
\vskip .6truein 
\centerline {F. del \'Aguila, $^{a,}$\footnote{e-mail: faguila@goliat.ugr.es}
             A. Culatti, $^{a,b,}$\footnote{e-mail:  culatti@mvxpd5.pd.infn.it}
             R. Mu\~noz Tapia, $^{a,}$\footnote{e-mail: rmt@ugr.es} and
             M. P\'erez-Victoria $^{a,c,}$\footnote{e-mail: mpv@ugr.es}}  
\end{center} 
\vspace{.3cm} 
\leftline 
{$^{a}$\UGR} 
\vspace{.1cm} 
\leftline
{$^{b}$\UP} 
\vspace{.1cm} 
\leftline
{$^{c}$\UKA} 

\vspace{1.5cm} 
 
\centerline{\bf Abstract} 
\medskip 

We present a procedure of  differential renormalization
at the one loop level which avoids introducing 
unnecessary renormalization constants and 
automatically preserves abelian gauge invariance. 
The amplitudes are expressed in terms of a basis of
singular functions. The local terms appearing in the
renormalization of these functions are determined
by requiring consistency with the propagator
equation.
Previous results in abelian theories, with and
without supersymmetry, are discussed in this context.  

\newpage
\pagestyle{plain}
\QQa
%\secteqno
\setcounter{footnote}{0}

%%%%%%%%%%%%%%%%%%%%%%%%
%    Intro             %
%%%%%%%%%%%%%%%%%%%%%%%%

Differential regularization and renormalization (DR)~\cite{FJL} 
was introduced as a renormalization method in coordinate space 
compatible with gauge and chiral symmetry. In a series of 
papers this method has been further
developed~\cite{counterterm,systematic,massiveDR} and successfully 
applied to different 
theories~[5--13]\footnote{Different versions of differential
renormalization can be found in~\cite{Smirnov,Schnetz}.}. 
\nocite{Susy,low,nonabelian,QED,chiral,anyon,curved,g2,
coreanos}
However, it might be considered unsatisfactory
the fact that Ward identities among renormalized Green functions 
are only satisfied when the different
renormalization scales are conveniently adjusted. 
Instead, one would like that the gauge symmetry were 
automatically preserved, as occurs in dimensional regularization 
and renormalization~\cite{dimreg}.  

In this letter we present  a procedure to constrain
the scales in DR at one loop while preserving
abelian gauge invariance. This is done in two steps. First, each diagram
is written in terms of a set of independent functions (with different
number of propagators and/or different tensor structure). Second, the 
singular functions of this set are renormalized in a way which does not 
depend on the diagram where
they appear. The local terms are fixed by the requirement that 
DR be compatible with the  
equation defining the propagator in the space of distributions. 
The propagator equation also allows to treat 
tadpole diagrams with
the usual DR
rules. In this manner one obtains renormalized Green functions
which depend on just one arbitrary constant (the renormalization group
scale) and, as we shall see, fulfil Ward identities in abelian
gauge theories. The non-abelian case will be studied elsewhere.

After describing the  method, we discuss the renormalization 
of the one-loop vacuum polarization in massive scalar QED, which is 
the simplest example requiring all the ingredients of the 
constrained procedure. The complete one-loop renormalization of this
theory will be presented in Ref.~\cite{SQED}. Then we review the
one-loop Ward identities of massive QED~\cite{massiveDR} 
and massless QED in an arbitrary
gauge~\cite{QED}, the corresponding ABJ anomaly~\cite{FJL,QED} 
and the evaluation of $(g-2)_l$ in supergravity, where supersymmetry
is also preserved~\cite{g2}.

%%%%%%%%%%%%%%%%%%%%%%%%%
%     METHOD            %
%%%%%%%%%%%%%%%%%%%%%%%%%

DR renormalizes diagrams by replacing singular expressions by
derivatives of well-behaved distributions ({\em differential reduction}). 
These derivatives are
understood in the sense of distribution theory, \ie, they are
prescribed to act formally by parts on test functions ({\em formal 
integration by parts}). In practice, to carry out this programme one
has to manipulate  singular expressions. 
This gives rise to ambiguities which are usually taken care of by 
keeping arbitrary renormalization scales for different diagrams
(or pieces of diagrams). The scales are adjusted at the end to enforce
the Ward identities (which is equivalent to the addition of finite
counterterms).
In this letter we show that four rules are sufficient to formally
manipulate and renormalize the singular expressions, avoiding the 
introduction of unnecessary scales. The resulting renormalized
amplitudes automatically satisfy the Ward identities. These rules
are summarised as follows:
\begin{enumerate}
  \item Differential reduction, 
        \label{R1} where we distinguish two cases:
     \begin{enumerate}
        \item Functions with singular behaviour worse than 
              $x^{-4}$ are reduced to derivatives of `logarithmically'
              singular functions without introducing extra dimensionful
              constants~\cite{Nuria}. For example, 
\be
  \frac{1}{x^6} = \frac{1}{8} \Box \frac{1}{x^4}~~.
\ee
              \label{R1a}
       \item Logarithmically singular functions are written as
              derivatives of regular functions. At one loop we
              have the usual DR identity~\cite{FJL}
\be
  \frac{1}{x^4} = -\frac{1}{4}  \Box \frac{\log x^2 M^2}{x^2}~~,
\ee 
              which introduces a unique dimensionful constant (the
              renormalization group scale). \label{R1b}
      \end{enumerate}   
  \item Formal integration by parts. \label{R2} In particular,
\be
  [\d F]^R = \d F^R ~,
\ee
        where $F$ is an arbitrary function and $R$ stands
        for renormalized.
  \item {\em Delta function renormalization rule}: \label{R3}
\be
  [F(x,x_1,...,x_n) \delta(x-y)]^R =  [F(x,x_1,...,x_n)]^R
  \delta(x-y)~.
\ee 
  \item The general validity of the {\em propagator equation}:
        \label{R4}
\be 
  F(x,x_1,...,x_n) (\Box^x - m^2) \propm(x) = 
  F(x,x_1,...,x_n) (- \delta(x)) ~,
  \label{masspropeq}
\ee
        where $\propm(x) = \frac{1}{4\pi^2} \frac{m K_1(mx)}{x}$ and
        $K_1$ is a modified Bessel function~\cite{Abramowitz}.
        This is a valid mathematical identity between tempered
        distributions if F is well-behaved enough. 
        This rule formally extends its range of applicability 
        to an arbitrary function.
\end{enumerate}
The last rule will prove essential in our procedure. 
For instance, the
`engineering' tensor decomposition into trace and traceless parts 
is not compatible with it and will be modified by the addition of 
a finite local term.

To evaluate the diagrams we express them in terms of a set of 
{\em basic functions} using only algebraic manipulations (including
four-dimensional Dirac algebra) and the Leibnitz rule for derivatives. 
To one loop these functions can be classified according to their number
of propagators and their derivative structure. For the examples we discuss
we need functions with one, two and three propagators: 
\bea
  \A  & = & \prop(x) \delta(x)~, \label{A} \\
  \B[\O] & = & \prop(x) \O^x \prop(x) ~, \label{B}\\
  \T[\O] & = & \prop(x) \prop(y) \O^x \prop(x-y)~, \label{T} 
\eea
where $\prop(x)  =  \frac{1}{4\pi^2} \frac{1}{x^2}$ is
the massless propagator
and $\O$ is a differential operator.
In massive theories the following basic functions
are also required:
\bea
  \bar{\A} & = & \bar{\prop}(x) \delta(x) ~, \label{Ab}\\
  \bar{\B}[\O] & = & \prop(x) \O^x \bar{\prop}(x) ~, \label{Bb}\\
  \bar{\T}[\O] & = & \prop(x) \prop(y) \O^x \bar{\prop}(x-y)~,
  \label{Tb}
\eea 
where $\bar{\prop}(x)  =  \frac{1}{4} \frac{1}{4\pi^2} 
\log x^2 m^2$ appears when the massive propagator is expanded in
the mass~$m$. Such expansion allows to properly separate pieces with
different degree of singularity. The same type of functions
also appear in massless theories if the photon propagator
is written in a general Lorentz gauge. 
Note that $\A$, $\bar{\A}$ and $\B$ functions are singular,
and
$\bar{\B}$ and $\T$ ($\bar{\T}$) are singular for 
$n \geq 2$ ($n \geq 4$), where $n$ is the order of the
differential operator $\O$.

We renormalize these basic functions using systematically
rules 1 to~4. For example, 
\bea
  \T[\Box]  & = & \prop(x) \prop(y) \Box^x \prop(x-y) \nn
            & = & - [\prop(x)]^2 \delta(x-y) \nn
            & = & - \B[1](x) \delta(x-y) ~; \\
  \T^R[\Box] & = & - \B^R[1](x) \delta(x-y) \nn
            & = & \frac{1}{4} \frac{1}{(4\pi^2)^2} \Box^x
                  \frac{\log x^2 M^2}{x^2} \delta(x-y) ~. 
\eea
Our main observation is that the propagator
equation (rule~\ref{R4}) can be further used to relate 
the different basic functions. 
Thus, by requiring that their renormalization 
be compatible with these relations, we shall completely 
fix the finite local terms (or scales) which appear in the 
differentially renormalized functions. 
Let us illustrate how to do this with one example, the renormalization 
of the basic function 
$\T[\d_\mu \d_\nu]$. 
Using rule~\ref{R4} for a massless propagator,
\be
  F \Box \prop(x) = - F \delta(x)~,
\label{mlesspropeq}
\ee
we can write
\bea 
  \B[\d_\mu](x) \delta(y)& = & - \d_\mu^x \Box^y \T[1]
  + \Box^y \T[\d_\mu] - 2 \d_\mu^x \d_\sigma^y \T[\d_\sigma]
  - \d_\mu^x \T[\Box] \nn
  && \mbox{}+ 2 \d_\sigma^y \T[\d_\mu\d_\sigma]
  + \T[\d_\mu \Box]~.
\label{relation}
\eea
The last basic function on the right-hand side can 
be easily reduced to
\be 
  \T[\d_\mu \Box] = \frac{1}{2} (\d_\mu^x - \d_\mu^y) \T[\Box]~,
\ee
where rules \ref{R3} and~\ref{R4} have been used.
Now, we decompose the basic function $\T[\d_\mu\d_\nu]$ into trace
and traceless parts, adding an arbitrary (for the moment) local
term to take into account the possible ambiguity introduced by
this operation: 
\be
  \T[\d_\mu\d_\nu] = \frac{1}{4} \delta_{\mu\nu} \T[\Box]
                    + \T[\d_\mu\d_\nu - \frac{1}{4} \delta_{\mu\nu}
                    \Box]
		    + \frac{1}{64 \pi^2} \, b \, \delta(x) \delta(y)
		    \delta_{\mu\nu} ~.
\label{trace}
\ee	
The traceless part is finite because of the tensor structure and is
not further renormalized. 
$\T[1]$ and $\T[\d_\mu]$ are also finite. 
On the other hand, the left-hand side of Eq.~(\ref{relation}) is
\bea
  \B[\d_\mu](x) \delta(y) & = & 
  \prop(x) \d_\mu^x \prop(x) \delta(y)  \nn  
  & = & \frac{1}{2} \d_\mu^x \prop(x)^2 \delta(y) \nn 
  & = & \frac{1}{2} \d_\mu^x \B[1](x) \delta(y) ~.
\eea  
So, Eq.~(\ref{relation})  reads for renormalized functions 
\bea
  \lefteqn{\frac{1}{2} \d_\mu^x \B^R[1](x) \delta(y)  =}   \nn  
  && - \d_\mu^x \Box^y \T[1]
     + \Box^y \T[\d_\mu] - 2 \d_\mu^x \d_\sigma^y \T[\d_\sigma]
     -\frac{1}{2} \d_\mu^x \T^R[\Box] + 2 \d_\sigma^y 
     \T[\d_\mu\d_\sigma - \frac{1}{4} \delta_{\mu\sigma}\Box] \nn
  && \mbox{} + \frac{1}{8} \frac{1}{4\pi^2} \, b \, \d_\mu^y 
     (\delta(x) \delta(y)) ~.
\eea
Since both members of this equation are finite,
we can integrate on $x$ using the integration
by parts prescription\footnote{Techniques for performing
this sort of integrals can be found in Appendix~A of Ref.~\cite{FJL}
and Appendix~C of Ref.~\cite{g2}.}:
\bea
  0 & = & \int \mathrm{d} x \,(\Box^y \T[\d_\mu] + 
      2 \d_\sigma^y \T[\d_\mu\d_\sigma - \frac{1}{4} 
      \delta_{\mu\sigma} \Box] )
      + \frac{1}{8} \frac{1}{4\pi^2} \, b \, \d_\mu^y \delta(y) \nn  
      & = & \frac{1}{16\pi^2} (\frac{1}{4} + \frac{1}{2} b) \,
      \d_\mu^y \delta(y) ~.
\label{intrel}
\eea
Then this equation fixes
\be
  b = - \frac{1}{2} ~.
\ee
Note that the engineering trace-traceless decomposition, commonly
used in the literature of DR, is not
compatible with the propagator equation in the case of logarithmic
singularities.
Hence, 
contraction of indexes does not commute with renormalization. 
Generically one must simplify all the tensor and Dirac structure before 
identifying the basic functions to be renormalized.
The addition of the local term is equivalent to using a different 
mass scale $M'$ in the renormalization of the
$\T[\Box]$ coming from the trace-traceless decomposition, and then 
fixing $\log \frac{M^2}{M'^2} = b = - \frac{1}{2}$. We prefer, however,
to use the language of local terms to avoid confusion with
the usual {\it ad hoc} adjustment of renormalization scales.

With the same technique one can  determine the renormalization
of all the basic functions. In general, besides the massless
propagator equation, Eq.~(\ref{mlesspropeq}), one needs
\be
  F \Box \bar{\prop}(x)  =   F \prop(x) ~.
\label{modpropeq}
\ee
Both Eq.~(\ref{mlesspropeq}) and Eq.~(\ref{modpropeq})
are  a consequence of  the massive propagator 
equation, Eq.~(\ref{masspropeq}), and equivalent to
the equation for the photon propagator
in a general gauge,
\be 
  F \,(\delta_{\mu\nu} \Box - (1-\frac{1}{a})\d_\mu\d_\nu) \, 
  \prop_{\nu\rho}(x) = - F \, \delta_{\mu\rho} \delta(x)~,
\label{apropeq}
\ee
where $\prop_{\mu\nu}(x) = 
1/16\pi^2  (\delta_{\mu\nu} \Box + (a-1) \d_\mu \d_\nu)  
\log x^2 \mu^2$.

The propagator equation can be further employed to `separate' the 
tadpole functions into two-point functions, which can then 
be treated
with the usual DR prescriptions:
\bea
  \A & = & \prop(x) \delta(x) \nn 
     & = & -\prop(x) \Box \prop(x) \nn  
     & = & - \B[\Box] ~.
\label{tadpole}
\eea

In Table~\ref{table1} we gather the renormalized expressions
of the basic functions required in the applications below.
%%%%%%%%%%%%%% TABLE 1 %%%%%%%%%%%%%%%%%%%
\begin{table}
\begin{center}
\begin{math}
\begin{array}{rcl}
\hline \\
  \A^R & = & 0  \\ \\
  \bar{\A}^R & = & -\Box \bar{\B}[1] + 2 \d_\sigma
    \bar{\B}[\d_\sigma] + \frac{1}{4} \frac{1}{(4\pi^2)^2}
    \Box \frac{\log x^2 M^2}{x^2}   \\ \\
  \B^R[1] & = & - \frac{1}{4} \frac{1}{(4\pi^2)^2} 
    \Box \frac{\log x^2 M^2}{x^2}  \\ \\
  \B^R[\Box] & = & 0  \\ \\
  \B^R[\d_\mu\d_\nu] & = & - \frac{1}{12}
    \frac{1}{(4\pi^2)^2} (\d_\mu\d_\nu - \frac{1}{4}
    \delta_{\mu\nu} \Box)
    \Box \frac{\log x^2 M^2}{x^2}\\ \\
    && \mbox{} +\frac{1}{18}\frac{\pi^2}{(4\pi^2)^2 }
    (\d_\mu\d_\nu-\delta_{\mu\nu} \Box) \delta(x)  \\ \\
  \bar{\B}^R[\Box] & = & \B^R[1]   \\ \\
  \bar{\B}^R[\d_\mu\d_\nu] & = &
    - \frac{1}{16} \frac{1}{(4\pi^2)^2} (
    \Box \frac{\log x^2 M^2}{x^2} \delta_{\mu\nu} 
    + 2 \d_\mu\d_\nu  \frac{1}{x^2})   \\ \\
  \T^R[\Box] & = & \frac{1}{4} \frac{1}{(4\pi^2)^2} 
    \Box \frac{\log x^2 M^2}{x^2} \delta(x-y)  \\ \\
  \T^R[\d_\mu\d_\nu] & = &
    (\frac{1}{16} \frac{1}{(4\pi^2)^2} 
    \Box \frac{\log x^2 M^2}{x^2} \delta(x-y) \\
    && \mbox{} -  \frac{1}{32} \frac{1}{4\pi^2} \delta(x) \delta(y) ) 
    \delta_{\mu\nu} 
    + \T[\d_\mu\d_\nu - \frac{1}{4} 
                     \delta_{\mu\nu}\Box] \\ \\
   \bar{\T}^R[\Box\Box] & = & \T^R[\Box]  \\ \\
   \bar{\T}^R[\Box\d_\mu\d_\nu] & = & 
     \T^R[\d_\mu\d_\nu]  \\ \\
\hline
\end{array} 
\end{math}
\end{center}
\caption{Renormalized expressions of basic functions.
         \label{table1}}
\end{table}
%%%%%%%%%%%%%%%%%%%%%%%%%%%%%%%%%%%%%%%%%%%%%%%%%%%
In massive theories it is usually more convenient to
work with compact expressions involving modified 
Bessel functions~\cite{massiveDR}. The 
corresponding DR identities can be obtained
by expanding the propagators in the mass parameter, 
using Table~\ref{table1} and resumming the result. 
In practice, one uses recurrence relations among
Bessel functions (see Appendix~C of Ref.~\cite{g2})
and then adds the necessary local terms  to agree with
Table~\ref{table1}. Table~\ref{table2} collects the
massive renormalization identities used in this paper.
%%%%%%%%%%%%%%% TABLE 2 %%%%%%%%%%%%%%%%%%%%%%%%%%%
\begin{table}
\begin{center}
\begin{math}
\begin{array}{rcl}
\hline \\
  \A_m^R & = & \frac{1}{(4\pi^2)^2} \pi^2  m^2 
    (1-\log \frac{\bar{M}^2}{m^2}) \delta(x)  \\ \\
  \B_m^R[1] & = & \frac{1}{(4\pi^2)^2} \{
    \frac{1}{2}(\Box-4m^2) \frac{m K_0(mx) K_1(mx)}{x}
    + \pi^2 \log \frac{\bar{M}^2}{m^2} \delta(x) \}  \\ \\
  \B_m^R[\Box] & = & \frac{1}{(4\pi^2)^2} m^2 \{
    \frac{1}{2}(\Box-4m^2) \frac{m K_0(mx) K_1(mx)}{x} \\
    && \mbox{} +  \pi^2  (2\log \frac{\bar{M}^2}{m^2} -1) 
    \delta(x) \}  \\ \\ 
  \B_m^R[\d_\mu\d_\nu] & = &  \frac{1}{(4\pi^2)^2}  \{
    \frac{1}{6} \d_\mu\d_\nu [
    (\Box-4m^2) ( \frac{m K_0(mx) K_1(mx)}{x} 
    + \frac{1}{4} m^2 (K_0^2(mx)-K_1^2(mx))) \\  
    && \mbox{} + 2\pi^2 ( \log \frac{\bar{M}^2}{m^2} -\frac{1}{3} ) 
    \delta(x) ] 
    \\
    && \mbox{} - \frac{1}{24} \delta_{\mu\nu} [
    (\Box-4m^2)(\Box-4m^2) \frac{m K_0(mx) K_1(mx)}{x} \\
    && \mbox{} + 2\pi^2 (\log \frac{\bar{M}^2}{m^2}+\frac{2}{3}) \Box \delta(x)
    - 4\pi^2 m^2 (1 + 3 \log \frac{\bar{M}^2}{m^2}) \delta(x) ]
    \}  \\ \\
\hline
\end{array}
\end{math}
\end{center}
\caption{Renormalized expressions of massive
         basic functions, \label{table2}
         where $\A_m  =  \propm(x) \delta(x)$ and
         $\B_m[\O]  = \propm(x) \O^x \propm(x)$.}
\end{table}
%%%%%%%%%%%%%%%%%%%%%%%%%%%%%%%%%%%%%%%%%%%%%%%%

%%%%%%%%%%%%%%%%%%%%%%%%%%%%%%%%%%%%
%    EXAMPLES                      %
%%%%%%%%%%%%%%%%%%%%%%%%%%%%%%%%%%%%

At this point any one-loop diagram can be renormalized:
one just has to use the 
renormalized basic functions listed in the Tables. 

As a simple example which contains all the
ingredients of the constrained procedure let us consider
in detail the vacuum polarization in massive scalar 
QED. 
The contributing diagrams are depicted in Fig.~\ref{fig}. 
Using the Feynman rules given in Ref.~\cite{g2}
one gets
\bea
  \Pi_{\mu\nu}^{(1)}(x) & = & 
  - e^2 \propm(x) 
  \stackrel{\leftrightarrow}{\d}_\mu
  \stackrel{\leftrightarrow}{\d}_\nu 
  \propm(x) ~, \\
  \Pi_{\mu\nu}^{(2)}(x) & = &  
  -2 e^2 \delta_{\mu\nu} \propm(x) \delta(x) ~,
\eea 
%%%%%%%%%%%%%%%%%%%%%%%%%%%%%%%%%%%
%    FIGURE                       %
%%%%%%%%%%%%%%%%%%%%%%%%%%%%%%%%%%%
%%%%%%%%%%%%%%%%%%%%%%%%%%%%%%%%%%%
\begin{figure}
\begin{tabular}{cc}
%%%%%%%%%%%%%%%%%%
%   Figure 1     %
%%%%%%%%%%%%%%%%%%
\mbox{\beginpicture
\setcoordinatesystem units <1.04987cm,1.04987cm>
\unitlength=1.04987cm
\linethickness=0.5pt
\setplotsymbol ({\makebox(0,0)[l]{\tencirc\symbol{'160}}})
\setlinear
%
% Fig ELLIPSE
%
\setdashes < 0.1270cm>
\ellipticalarc axes ratio  0.620:0.620  360 degrees 
	from  3.715 24.765 center at  3.095 24.765
\setsolid
%
% Fig INTERPOLATED PT SPLINE
%
\plot  0.953 24.765 	 0.975 24.855
	 0.997 24.914
	 1.048 24.955
	 1.107 24.896
	 1.126 24.835
	 1.143 24.765
	 1.160 24.695
	 1.179 24.634
	 1.238 24.575
	 1.297 24.634
	 1.317 24.695
	 1.334 24.765
	 1.350 24.835
	 1.370 24.896
	 1.429 24.955
	 1.488 24.896
	 1.507 24.835
	 1.524 24.765
	 1.541 24.695
	 1.560 24.634
	 1.619 24.575
	 1.678 24.634
	 1.698 24.695
	 1.715 24.765
	 1.731 24.835
	 1.751 24.896
	 1.810 24.955
	 1.869 24.896
	 1.888 24.835
	 1.905 24.765
	 1.922 24.695
	 1.941 24.634
	 2.000 24.575
	 2.059 24.634
	 2.079 24.695
	 2.095 24.765
	 2.112 24.835
	 2.132 24.896
	 2.191 24.955
	 2.250 24.896
	 2.269 24.835
	 2.286 24.765
	 2.303 24.695
	 2.322 24.634
	 2.381 24.575
	 2.432 24.616
	 2.454 24.675
	 2.477 24.765
	/
%\setplotsymbol ({\thinlinefont .})
%
% Fig INTERPOLATED PT SPLINE
%
\plot  3.715 24.765 	 3.739 24.676
	 3.761 24.617
	 3.810 24.575
	 3.870 24.634
	 3.889 24.695
	 3.906 24.765
	 3.923 24.835
	 3.942 24.896
	 4.000 24.955
	 4.060 24.895
	 4.080 24.835
	 4.097 24.765
	 4.113 24.695
	 4.132 24.635
	 4.191 24.575
	 4.251 24.634
	 4.270 24.695
	 4.287 24.765
	 4.304 24.835
	 4.323 24.896
	 4.381 24.955
	 4.441 24.895
	 4.461 24.835
	 4.478 24.765
	 4.494 24.695
	 4.513 24.635
	 4.572 24.575
	 4.632 24.634
	 4.651 24.695
	 4.668 24.765
	 4.685 24.835
	 4.704 24.896
	 4.763 24.955
	 4.822 24.895
	 4.842 24.835
	 4.859 24.765
	 4.875 24.695
	 4.894 24.635
	 4.953 24.575
	 5.013 24.634
	 5.032 24.695
	 5.049 24.765
	 5.066 24.835
	 5.085 24.896
	 5.143 24.955
	 5.194 24.913
	 5.216 24.855
	 5.239 24.765
	/
\linethickness=0pt
\putrectangle corners at  0. 26.0 and  6.5 23.5
\endpicture}&
%%%%%%%%%%%%%%%%%%
%   Figure 2     %
%%%%%%%%%%%%%%%%%%
\mbox{\beginpicture
\setcoordinatesystem units <1.04987cm,1.04987cm>
\unitlength=1.04987cm
\linethickness=0.5pt
\setplotsymbol ({\makebox(0,0)[l]{\tencirc\symbol{'160}}})
\setshadesymbol ({\thinlinefont .})
\setlinear
%
% Fig ELLIPSE
%
\setdashes < 0.1270cm>
\ellipticalarc axes ratio  0.667:0.904  360 degrees 
	from  3.619 25.576 center at  2.953 25.576
\setsolid
%
% Fig INTERPOLATED PT SPLINE
%
\plot  2.477 24.479 	 2.499 24.569
	 2.521 24.628
	 2.572 24.670
	 2.631 24.610
	 2.650 24.549
	 2.667 24.479
	 2.684 24.409
	 2.703 24.348
	 2.762 24.289
	 2.821 24.348
	 2.841 24.409
	 2.857 24.479
	 2.874 24.549
	 2.894 24.610
	 2.953 24.670
	 3.012 24.610
	 3.031 24.549
	 3.048 24.479
	 3.065 24.409
	 3.084 24.348
	 3.143 24.289
	 3.202 24.348
	 3.222 24.409
	 3.239 24.479
	 3.255 24.549
	 3.275 24.610
	 3.334 24.670
	 3.393 24.610
	 3.412 24.549
	 3.429 24.479
	 3.446 24.409
	 3.465 24.348
	 3.524 24.289
	 3.583 24.348
	 3.603 24.409
	 3.620 24.479
	 3.636 24.549
	 3.656 24.610
	 3.715 24.670
	 3.774 24.610
	 3.793 24.549
	 3.810 24.479
	 3.827 24.409
	 3.846 24.348
	 3.905 24.289
	 3.956 24.330
	 3.978 24.389
	 4.000 24.479
	/
%
% Fig INTERPOLATED PT SPLINE
%
\plot  4.000 24.479 	 4.023 24.569
	 4.045 24.628
	 4.096 24.670
	 4.155 24.610
	 4.174 24.549
	 4.191 24.479
	 4.208 24.409
	 4.227 24.348
	 4.286 24.289
	 4.345 24.348
	 4.365 24.409
	 4.381 24.479
	 4.398 24.549
	 4.418 24.610
	 4.477 24.670
	 4.536 24.610
	 4.555 24.549
	 4.572 24.479
	 4.589 24.409
	 4.608 24.348
	 4.667 24.289
	 4.726 24.348
	 4.746 24.409
	 4.763 24.479
	 4.779 24.549
	 4.799 24.610
	 4.858 24.670
	 4.917 24.610
	 4.936 24.549
	 4.953 24.479
	 4.970 24.409
	 4.989 24.348
	 5.048 24.289
	 5.099 24.330
	 5.121 24.389
	 5.143 24.479
	/
%
% Fig INTERPOLATED PT SPLINE
%
\plot  0.953 24.479 	 0.975 24.569
	 0.997 24.628
	 1.048 24.670
	 1.107 24.610
	 1.126 24.549
	 1.143 24.479
	 1.160 24.409
	 1.179 24.348
	 1.238 24.289
	 1.297 24.348
	 1.317 24.409
	 1.334 24.479
	 1.350 24.549
	 1.370 24.610
	 1.429 24.670
	 1.488 24.610
	 1.507 24.549
	 1.524 24.479
	 1.541 24.409
	 1.560 24.348
	 1.619 24.289
	 1.678 24.348
	 1.698 24.409
	 1.715 24.479
	 1.731 24.549
	 1.751 24.610
	 1.810 24.670
	 1.869 24.610
	 1.888 24.549
	 1.905 24.479
	 1.922 24.409
	 1.941 24.348
	 2.000 24.289
	 2.059 24.348
	 2.079 24.409
	 2.095 24.479
	 2.112 24.549
	 2.132 24.610
	 2.191 24.670
	 2.250 24.610
	 2.269 24.549
	 2.286 24.479
	 2.303 24.409
	 2.322 24.348
	 2.381 24.289
	 2.432 24.330
	 2.454 24.389
	 2.477 24.479
	/
\linethickness=0pt
\putrectangle corners at  -0.2 27.3 and 6.5 23.272
\endpicture} \\
Diagram 1 & Diagram 2
\end{tabular}
\caption{One-loop diagrams contributing to the 
vacuum polarization of scalar QED. \label{fig}}
\end{figure}
%%%%%%%%%%%%%%%%%%%%%%%%%%%%%%%%%%%

\noindent which expressed in terms of basic functions read
\bea
  \Pi_{\mu\nu}^{(1)}(x) & = & -e^2 \{ 4\B_m[\d_\mu\d_\nu]
    - \d_\mu\d_\nu B_m[1] \} ~, \\
  \Pi_{\mu\nu}^{(2)}(x) & = & -2 e^2 \delta_{\mu\nu} \A_m ~.
\eea   
Substituting the renormalized basic functions of 
Table~\ref{table2}, we obtain for each diagram
\bea
  \Pi_{\mu\nu}^{(1)\,R}(x) & = & - \frac{e^2}{(4\pi^2)^2} \{
    (\d_\mu\d_\nu - \delta_{\mu\nu} \Box) [ 
    \frac{1}{6} (\Box - 4m^2) (\frac{m K_0(mx) K_1(mx)}{x} \nn
    && \mbox{} + m^2 (K_0^2(mx)-K_1^2(mx))) + 
    \frac{1}{3} \pi^2 (\log \frac{\bar{M}^2}{m^2} - \frac{4}{3}) 
    \delta(x) ]  \nn
    & & \mbox{} + \delta_{\mu\nu} [ 2 \pi^2 m^2 
    (\log \frac{\bar{M}^2}{m^2} -1) \delta(x) ] \} ~, \\
  \Pi_{\mu\nu}^{(2)\,R}(x) & = &  \frac{e^2}{(4\pi^2)^2}
    2 \pi^2 m^2 (\log \frac{\bar{M}^2}{m^2} -1) 
    \delta_{\mu\nu} \delta(x) ~.
\eea        
The longitudinal terms cancel in the sum, yielding a 
transverse result:
\bea
  \Pi_{\mu\nu}^R(x) & = & - \frac{e^2}{(4\pi^2)^2} 
    (\d_\mu\d_\nu - \delta_{\mu\nu} \Box) [ 
    \frac{1}{6} (\Box - 4m^2) (\frac{m K_0(mx) K_1(mx)}{x} \nn
    && \mbox{} +  m^2 (K_0^2(mx)-K_1^2(mx))) + 
    \frac{1}{3} \pi^2 (\log \frac{\bar{M}^2}{m^2} - \frac{4}{3}) 
    \delta(x) ]  ~.
\eea

In the same  way, our method recovers previous
DR results in  abelian
gauge theories, without the need to impose Ward identities
{\it a posteriori}. Let us consider first the vacuum 
polarization in massive QED~\cite{massiveDR}. In terms of
basic functions it reads
\be
  \Pi_{\mu\nu}(x) = 4 e^2 \{ (m^2 \delta_{\mu\nu}
  + \frac{1}{2} \delta_{\mu\nu} \Box - \d_\mu\d_\nu) \,
  \B_m[1] + 2 \B_m[\d_\mu\d_\nu] - \delta_{\mu\nu} 
  \B_m[\Box] \} 
\ee
and the renormalized expression is
\bea
  \Pi_{\mu\nu}^R(x) & = & - \frac{4 e^2}{(4\pi^2)^2} 
  (\d_\mu\d_\nu - \delta_{\mu\nu} \Box) [
  \frac{1}{6} (\Box - 4m^2) (\frac{m K_0(mx) K_1(mx)}{x} \nn
  && \mbox{} - \frac{1}{2} m^2 (K_0^2(mx)-K_1^2(mx)))
  + \frac{1}{3} \pi^2 (\log \frac{\bar{M}^2}{m^2} + \frac{2}{3})
  \delta(x) ]  ~,
\eea
which is  again transverse.
 
In supersymmetric QED the vacuum polarization is the
sum of the spinor QED diagram and twice (two complex scalars 
for each Dirac spinor) the scalar QED diagrams. In terms
of basic functions this gives directly a
transverse result depending on 
one basic function only:
\be 
  \Pi_{\mu\nu}(x) = -2 e^2 (\d_\mu\d_\nu - \delta_{\mu\nu} \Box)
  \B_m[1]~.
\ee
In this case one could 
use  an engineering 
trace-traceless decomposition in renormalizing each diagram. The 
gauge-non-invariant terms  vanish
in the total sum due to supersymmetry cancellations. 
The complete result would thus be the 
same as the one obtained from the renormalized
functions of Table~\ref{table2}:
\bea
  \Pi_{\mu\nu}^R(x) & = & - \frac{ e^2}{(4\pi^2)^2} 
  (\d_\mu\d_\nu - \delta_{\mu\nu} \Box) [
  (\Box - 4m^2) \frac{m K_0(mx) K_1(mx)}{x} \nn
  && \mbox{} + 2 \pi^2 \log \frac{\bar{M}^2}{m^2} 
  \delta(x) ]  ~.
\eea

Next we consider the QED vertex Ward identity between the electron 
self-energy and the electron-electron-photon vertex in an arbitrary
Lorentz gauge, which was studied in Ref.~\cite{QED}. In this case 
and the next one the masses play no 
relevant role, as far as renormalization is concerned, 
so we consider massless electrons for simplicity.
In order to respect the Ward identity for both the $a$-dependent 
and $a$-independent pieces (where $a$ is the gauge parameter), the
authors of Ref.~\cite{QED} had to impose  two relations among 
different scales:
\bea
  & \log \frac{M_\Sigma}{M_V}  =  \frac{1}{4} ~, \\
  & \lambda  \equiv  \log \frac{M_\Sigma}{M_\Sigma'}  = 3 ~,
\eea
where $M_V$ and $M_\Sigma$ and $M_\Sigma'$ appear
in the vertex and in the two pieces of the electron
self-energy, respectively.
With the constrained method we find 
\bea
  \Sigma^R(x) & = & e^2 \{\frac{1}{4} \frac{1}{(4\pi^2)^2}
  \dsl \Box  \frac{\log x^2 M^2}{x^2}  \nn
  && \mbox{} + (a-1) [ \frac{1}{4} \frac{1}{(4\pi^2)^2} 
  \dsl \Box  \frac{\log x^2 M^2}{x^2} +
  \frac{1}{16\pi^2} \dsl \delta(x) ] \}
\eea 
for the electron self-energy, and
\bea
  V_\mu^R(x,y) & = & ie^3 \{  -2 \g_b \g_\mu \g_a (
  \d_a^x \d_b^y \bar{\T}[\Box] + \d_a^x \bar{\T}[\d_b \Box]
  - \d_b^y \bar{\T}[\d_a \Box])  
  + 4 \g_a \T[\d_a\d_\mu - 
  \frac{1}{4}\delta_{a\mu} \Box] \nn
  && \mbox{} - \frac{1}{4}
  \frac{1}{(4\pi^2)^2} \g_\mu \Box \frac{\log x^2 M^2}{x^2}
  \delta(x-y)
  - \frac{1}{8} \frac{1}{4 \pi^2} 
  \g_\mu \delta(x) \delta(y) \nn
  && \mbox{} + (a-1) [ \g_\rho \g_a \g_\mu \g_b \g_\sigma
  (\d_a^x \d_b^y \bar{\T}[\d_\rho\d_\sigma] +
  \d_a^x \bar{\T}[\d_b \d_\rho \d_\sigma] 
  - \d_b^y \bar{\T}[\d_a \d_\rho \d_\sigma])  \nn
  && \mbox{} -  \frac{1}{4}
  \frac{1}{(4\pi^2)^2} \g_\mu \Box 
  \frac{\log x^2 M^2}{x^2} \delta(x-y)] \}
\eea
for the vertex.
These renormalized amplitudes automatically
satisfy the Ward identity
\be 
  (\d_\mu^x + \d_\mu^y) V_\mu^R(x,y) =
  i e \Sigma^R(x-y) (\delta(x) - \delta(y)) ~,
\ee
as can be seen by integrating on $y$. 

The chiral triangle anomaly in QED was discussed 
in Refs.~\cite{FJL,QED}. The renormalized
triangle diagram depended on the relation between two
renormalization scales. By adequately choosing these
scales one could respect either vector current
or axial current conservation, but not both. Imposing
conservation of the vector current, the correct
value of the axial anomaly resulted.
In contrast, in the constrained DR method  
everything is determined and   
a non-ambiguous result is obtained:
\bea
  \T_{\mu\nu\lambda}^R(x,y) & = &
  i e^3 \{ -2 \mathrm{tr}
  (\g_5\g_\mu\g_\lambda\g_\nu\g_a\g_b\g_c)
  \d_c^x\d_a^y \T[\d_b] +
  16 (\epsilon_{\lambda\mu ab}\d_a^x\d_\nu^y
  - \epsilon_{\lambda\nu ab}\d_\mu^x\d_a^y)
  \T[\d_b]  \nn
  && \mbox{} + 16 \epsilon_{\lambda b \mu a}\d_a^x
  \T[\d_\nu\d_b - \frac{1}{4}\delta_{\nu b} \Box] 
  - 16 \epsilon_{\lambda \nu ba}\d_a^y 
  \T[\d_\mu\d_b - \frac{1}{4}\delta_{\mu b} \Box] \nn
  && \mbox{} + \frac{1}{8\pi^2} 
  \epsilon_{\mu\nu\lambda a} (\d_a^x-\d_a^y) 
  (\delta(x) \delta(y)) \}~,
\eea
where the index $\lambda$ corresponds to the axial
vertex. Then 
(see appendix B of Ref.~\cite{FJL}) 
\bea
  \d_\mu^x T_{\mu\nu\lambda}^R(x,y) & = & 0 ~, \\
  \d_\nu^y T_{\mu\nu\lambda}^R(x,y) & = & 0 ~, \\
  - (\d_\lambda^x + \d_\lambda^y) T_{\mu\nu\lambda}^R(x,y)
  &  =  & \frac{i e^3}{2\pi^2} \epsilon_{\mu\nu\lambda\rho}
  \d_\lambda^x \d_\rho^y (\delta(x) \delta(y))~,
\eea
so the vector Ward identities are directly 
preserved while the axial one is 
broken, giving the known result for the anomaly. 

Finally, let us comment briefly on 
the calculation of the
$(g-2)_l$ in unbroken supergravity  performed
in Ref.~\cite{g2}.  There, the symmetry to be preserved was
supersymmetry, which implies a vanishing anomalous 
magnetic moment~\cite{Ferrara}.
This result was obtained thanks to the use of 
the propagator equation to explicitly relate  diagrams
with different topology. Then, only one type of
singular basic function, $\T[\Box]$, appeared. 
Although  engineering trace-traceless
decompositions were performed at intermediate steps, this
did not affect the total sum because the extra
local terms cancel, as occurs in the vacuum polarization
in supersymmetric QED discussed above. 
Using the renormalized basic
functions in Tables~\ref{table1} and~\ref{table2},
$(g-2)_l$ also vanishes,
although the contribution of each diagram is different,
as is the total graviton contribution.

Summarizing, we have proposed a procedure of differential
renormalization to one loop which only introduces a single
renormalization scale. We have verified that the
renormalized amplitudes so obtained automatically
satisfy the Ward identities of abelian gauge symmetry in
known examples, that the chiral anomaly is
correctly treated, and that supersymmetry is preserved
in a relatively complex calculation.
In practice, one just needs to use the renormalized
functions of Tables~\ref{table1} and~\ref{table2}.

In principle, the method could be generalized to
higher loops. However this is not straightforward.
New more complicated functions emerge, which could
be tackled with the systematic method of
Ref.~\cite{systematic}. Still, one should  
properly constrain
the local terms using rules 1 to 4 or
some consistent extension of them.

We thank  J. Collins and R. Stora for discussions. 
This work has been supported by CICYT, contract
number AEN96-1672, and by Junta de Andaluc\'{\i}a, 
FQM101.
RMT and MPV thank MEC for financial support.

%%%%%%%%%%%%%%%%%%%%%%%%%
%  Bibliografia         %
%%%%%%%%%%%%%%%%%%%%%%%%%

\end{document}